\documentclass[useAMS,usenatbib]{mn2e}

\usepackage{graphicx}
\usepackage{txfonts}
\usepackage{color}

\newcommand{\fig}[1]{Fig.~\ref{#1}}
\newcommand{\sect}[1]{Sect.~\ref{#1}}
\newcommand{\eq}[1]{Eq.~(\ref{#1})}
\def\spose#1{\hbox to 0pt{#1\hss}}
\def\gsim{\mathrel{\spose{\lower 3pt\hbox{$\mathchar"218$}}
          \raise 2.0pt\hbox{$\mathchar"13E$}}}
\def\lsim{\mathrel{\spose{\lower 3pt\hbox{$\mathchar"218$}}
          \raise 2.0pt\hbox{$\mathchar"13C$}}}

\title[Prompt-early afterglow connection in gamma-ray bursts]{The prompt-early afterglow connection in gamma-ray bursts: implications for 
the early afterglow physics}
\author[R. Hasco\"et et al.]{R. Hasco\"et$^{1}$\thanks{E-mail:
hascoet@astro.columbia.edu}, F. Daigne$^{2}$ and R. Mochkovitch$^{2}$\footnotemark[1]\\
$^{1}$Physics Department and Columbia Astrophysics Laboratory, Columbia University, 538 West 120th Street, New York, NY 10027, USA.\\
$^{2}$UPMC-CNRS, UMR7095, Institut d'Astrophysique de Paris, F-75014, Paris, France.}
\begin{document}

\date{Accepted **.**.**. Received **.**.**; in original form **.**.**}

\pagerange{\pageref{firstpage}--\pageref{lastpage}} \pubyear{2002}

\maketitle

\label{firstpage}

\begin{abstract}
The early X-ray afterglow of gamma-ray bursts revealed by \textit{Swift} carried many surprises. Following an initial steep
decay the light-curve often exhibits a plateau phase that can last up to several $10^4$ s, with in addition the
presence of flares in 50\% of the cases. We 
focus
in this paper on the plateau phase whose origin remains highly debated.
We confront several newly discovered correlations between prompt and afterglow quantities (isotropic emitted energy in
gamma-rays, luminosity and duration of the plateau) to 
several models proposed for the origin of plateaus
in order to check if they can account for these observed correlations.
We first show that the 
scenario of plateau formation by energy injection into the forward shock leads to an efficiency crisis for the prompt phase and
therefore study two possible alternatives: the first one still takes place within the framework of the standard forward shock
model but allows for a variation of the microphysics parameters to reduce the radiative efficiency at early times; in the second scenario
the early afterglow results from a long-lived reverse shock. Its shape then depends on the distribution of
energy as a function of Lorentz factor in the ejecta. In both cases, we first present simple 
analytical estimates of the plateau luminosity and duration and then compute detailed light curves. 
In the two considered scenarios we find that plateaus following the observed correlations can be obtained
under the condition that specific additional ingredients are included. In the forward shock scenario, the
preferred model supposes a wind external medium and a microphysics parameter $\epsilon_e$ that first
varies as $n^{-\nu}$ ($n$ being the external density),
 with $\nu\sim 1$ to get a flat plateau,
before staying constant below a critical density $n_0$.
To produce a plateau in the reverse shock scenario the ejecta must contain a tail of low Lorentz factor with a peak of energy deposition 
at $\Gamma \ga 10$. 
\end{abstract}



\maketitle
\section{Introduction}
Before the launch of the {\it Swift} satellite \citep{gehrels_2004} the afterglow was believed to be the best understood part of GRB physics,
being explained by the energy dissipated in the forward shock formed by the jet impacting the burst environment
\citep{meszaros_1997, sari_1998}. 
However, the many surprises of the early X-ray afterglow revealed by {\it Swift} - initial steep decay, plateau phase, flares - have 
considerably complicated the picture \citep{nousek_2006, obrien_2006}. 

Several mechanisms have been proposed to explain the plateau, the most popular being energy injection into the forward 
shock \citep{rees_1998, sari_2000, nousek_2006} resulting from a long-lasting activity of the central engine 
(which could be also responsible for the flares, \citealt{zhang_2006}) 
or from a wide distribution of Lorentz factors in the ejecta. 
Other possibilities include {\it (i)} direct emission from a magnetar (e.g. \citealt{rowlinson_2013}),
{\it (ii)} coasting of the external blastwave in a wind medium (e.g. \citealt{shen_2012}),
{\it (iii)} varying microphysics parameters \citep{granot_2006, ioka_2006}, 
{\it (iv)} reverse shock contribution \citep{genet_2007, uhm_2007}. 
In {\it (i)} the end of the plateau corresponds to the spindown time of the protomagnetar or its collapse to a blackhole.
Therefore this scenario is mostly promising to explain peculiar plateaus that are followed by a steep decay (temporal index $\sim -2$ or steeper),
while ``standard'' plateaus (followed by a temporal decay index $\sim -1.5$) are most likely of afterglow origin; {\it (ii)} requires 
the Lorentz factor of the ejecta to be at most a few tens 
(so that the coasting phase lasts long enough), which is in severe tension with the minimum Lorentz factor of the ejecta derived 
from the compactness constraint (e.g. \citealt{lithwick_2001, hascoet_2012}).
In the present work we focus on cases {\it (iii)} and {\it (iv)} in connection with the recent discovery of correlations
between prompt and afterglow quantities \citep{dainotti_2011, margutti_2013, dainotti_2013, grupe_2013}. We especially want to explore if these correlations can be satisfied by the models
and which kind of constraints do they impose.  

We first summarize in \sect{sect_obs} the observational results on the prompt-afterglow correlations and in \sect{sect_lateInject} we show that explaining the plateau 
by late energy injection into the forward shock leads to an ``efficiency crisis'' for the prompt phase. We then consider in \sect{sect_alternatives} the
possibility that the microphysics parameters in the forward shock vary during the early afterglow and in \sect{sect_sequence} we explore the alternative model where
the afterglow is made by the reverse shock. Our results are discussed in \sect{sect_conclusion}, which is also the conclusion.

\section{The prompt afterglow connection}
\label{sect_obs}

For events with a measured redshift and a well-defined plateau phase, 
quantities such as $t_{\rm P}$ -- duration of the plateau in the burst
rest frame, $L_{\rm P}$
-- luminosity at the end of the plateau or $E_{\rm X}$ -- energy released in X-rays during the plateau, can be measured together with 
the isotropic energy in gamma-rays of the prompt phase $E_{\gamma,{\rm iso}}$. 
From the samples recently analyzed by \citet{dainotti_2011}, \citet{margutti_2013} and \citet{dainotti_2013} some clear correlations appear between prompt and 
afterglow quantities.
The plateau luminosity $L_{\rm p}$ and energy $E_X$ increase with $E_{\gamma,{\rm iso}}$ and decrease for larger $t_{\rm p}$. Since an increase of $L_{\rm p}$ and $E_X$ with $E_{\gamma,{\rm iso}}$
could be expected, we also consider below the ratios $L_{\rm p}/E_{\gamma,{\rm iso}}$ and $E_X/E_{\gamma,{\rm iso}}$, which respectively decreases and barely evolves with increasing $t_{\rm p}$.

These prompt-afterglow correlations represent potentially important clues to understand the many surprises of the early afterglow. 
In the standard forward shock scenario (for a wide range of parameters) the X-ray flux depends on the energy injected into the shock and the microphysics, 
but not on the density of external medium.                                                                                                                                 
In the reverse shock scenario the shape of the early afterglow depends both on the density of the burst environment and on the distribution 
of energy in the ejecta that is crossed by the reverse shock.   
Below, we investigate under which conditions 
the observed correlations can be reproduced in the framework of these two scenarios.
  
\section{Making a plateau with late energy injection}
\label{sect_lateInject}

\begin{figure}
\begin{center}
\includegraphics[width=0.45\textwidth]{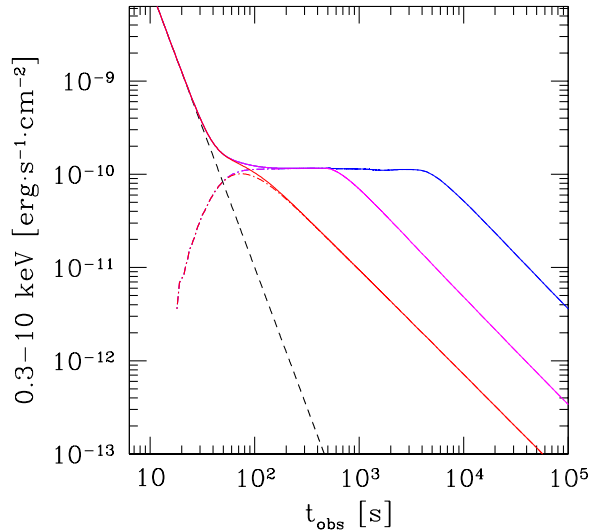} 
\end{center}
\caption{X-ray afterglow light curves from late energy injection into the forward shock. The initial energy in the shock is $E_0=10^{52}$ erg and the red, purple and 
blue light curves respectively correspond to a final energy being respectively 2, 10 and 100 times larger. The dashed line represents the continuation of the 
early steep decay that terminates the prompt emission, while the dashed-dotted line shows the forward shock emission only.   
A redshift $z=1$, a uniform external medium of density $n=10 \ {\rm cm^{-3}}$, and constant microphysics parameters $\epsilon_e = 0.1$ and $\epsilon_B = 0.01$ have been assumed.}
\label{fig_lateInject}
\end{figure}

Continuous energy injection into the forward shock \citep{rees_1998, sari_2000, nousek_2006} is commonly invoked to account for plateau formation. 
For the most extended plateaus
it however imposes to inject up to several hundreds times the energy that was initially present to power the prompt phase. 
This is illustrated in \fig{fig_lateInject} where we have plotted X-ray light curves all with the same initial injected energy $E_0=10^{52}$ erg
but where the final energy is 2, 10 or 100 times larger. It is only in this last case that a plateau
lasting several hours can be obtained. Energy injection into the forward shock can take place in two ways: either the source stays active 
during the whole duration of the plateau or it is short-lived but has produced a tail of low Lorentz factor material that is progressively
catching up, adding energy to the shock. We have considered this latter case to obtain Fig.1 (the source being active for 10 s) 
but the former one gives similar results.     

The huge amount of energy to be injected after the end of the prompt phase leads to an ``efficiency crisis'' for the prompt mechanism.
The measured gamma-ray efficiency is
\begin{equation}
f_{\gamma,{\rm mes}}={E_\gamma\over E_\gamma+E_{\rm fs}}
\end{equation}
where the energy in the forward shock, $E_{\rm fs}$, is estimated from multiwavelength fits of the afterglow typically after one day (i.e. after 
energy injection; see e.g. \citealt{zhang_2007}). However the true efficiency is
\begin{equation}
f_{\gamma,{\rm true}}={E_\gamma\over E_\gamma+E_{{\rm fs},0}}={1\over 1+{1\over k}\left({1\over f_{\gamma,{\rm mes}}}-1\right)}
\end{equation} 
where $E_{{\rm fs},0}$ is the energy initially present in the forward shock and  
$k=E_{\rm fs}/E_{{\rm fs},0}\gg 1$. With for example $f_{\gamma,{\rm mes}}=0.1$, the true efficiency is $f_{\gamma,{\rm true}}=0.53$ 
for $k=10$ and $0.92$ for $k=100$.
These values of $f_{\gamma,{\rm true}}$ seems unreachable for any of the proposed prompt mechanisms: the efficiency of internal shocks can barely reach 10\% 
(e.g. \citealt{rees_1994, kobayashi_1997, daigne_1998}) while
that of comptonized photosphere (e.g. \citealt{ress_2005, beloborodov_2010}) or reconnection (e.g. \citealt{spruit_2001, drenkhahn_2002}) models 
is more uncertain but certainly cannot exceed 50\%. 

\section{Making a plateau avoiding an energy crisis}
\label{sect_alternatives}

\subsection{Forward shock scenario}
\label{subsect_fsPrinciple}

The standard forward shock scenario can successfully account for the afterglow evolution after 
about one day but fails to reproduce the plateau phase. A backwards extrapolation of the late afterglow flux lies above the plateau, which might
therefore be interpreted as the indication that some normally expected radiation is ``missing''. This can be the case if the radiative efficiency 
of the forward shock during the early afterglow is smaller than assumed by the simplest version of the standard model. The most obvious way to reduce the efficiency 
is to relax the assumption that the microphysics parameters stay constant throughout the whole afterglow evolution \citep{granot_2006, ioka_2006}. 

For both a uniform and a wind external medium the afterglow X-ray flux behaves as \citep{panaitescu_2000}
\begin{equation}
\label{eq_fx}
F_{\rm X}\propto E^{p+2\over 4}\epsilon_e^{p-1}\epsilon_B^{p-2\over 4}t^{-{3p-2\over 4}}
\end{equation}    
where $E$ is the burst isotropic energy, $\epsilon_e$ and $\epsilon_B$ the microphysics parameters and $p$ the power-law index of the
accelerated electron spectrum. 
\eq{eq_fx} is valid as long as the X-ray frequency is larger than both the injection and cooling frequencies, which is generally the case. 

With $2<p<3$ the dependence on $\epsilon_B$ is weak so that in practice 
only playing with $\epsilon_e$ can really affect the flux evolution. A priori $\epsilon_e$ can be a function of the shock Lorentz factor, the density of
the external medium (in the case of a stellar wind) or both. 
The stellar wind case is of special interest if we make 
the simple assumption that, below a critical
density $n_0$, $\epsilon_e$ is constant while $\epsilon_e\propto n^{-\nu}$ (with $\nu>0$) for $n>n_0$. 
Since the density seen by the forward shock is given by 
\begin{equation}
\label{eq_ntlab}
n(t) \simeq \frac{4 \pi c}{m_p}\frac{A^2}{E \  t} \simeq 5.6\,10^2 A_*^2 E_{53}^{-1} t_3^{-1}$ cm$^{-3}
\end{equation}
where $t$ is the (redshift-corrected) observer time and $A_*$ is the wind density normalization 
($\rho(r) = A /R^2$ with $A=5\times 10^{11} A_* \ \mathrm{g\ cm^{-1}}$) the transition at $n_0$, which marks the end of the plateau, takes place at
\begin{equation}
\label{eq_tp_n0}
t_{\rm p} \approx 5.6\,10^5 A_*^2 n_0^{-1} f_{\gamma} E_{\gamma,53}^{-1}\ \ {\rm s}
\end{equation}  
where $f_{\gamma}$ is the gamma-ray efficiency of the prompt phase and $E_{\gamma,53}$ is the isotropic gamma-ray energy release. 
Then, if the product $A_*^2 n_0^{-1} f_{\gamma}$ typically stays in the range 
$3\times10^{-4} - 3\times10^{-2}$
the resulting [$t_{\rm p},E_{{\gamma},{\rm iso}}$] sequence can accommodate most of the bursts in the \citet{margutti_2013} sample 
(see \fig{fig_correlations}). 

A flat plateau is expected for 
\begin{equation} 
\nu=\nu_0={3p-2\over 4(p-1)}=1-{p-2\over 4(p-1)}\approx 1
\end{equation}
while for $\nu<\nu_0$ (resp. $\nu>\nu_0$) the plateau flux is decreasing (resp. rising) with time. 

With $\epsilon_e\propto n^{-1}$ and from \eq{eq_fx},
a flat plateau extending over two decades in time requires an increase of $\epsilon_e$ by a factor of about $100$ from the
beginning to the end of the plateau. It is beyond the scope of this paper to decide if this is indeed possible 
but it is remarkable
that acting on one single parameter can lead to the formation of a plateau that also satisfies 
the observed prompt-afterglow correlations (see \S\ref{fs_scenario}).

The other possibility where $\epsilon_e$ depends on the Lorentz factor does not yield satisfactory
results. Assuming that the transition from a varying to a constant $\epsilon_e$ takes place at a fixed $\Gamma$, the deceleration 
laws of the blast wave
\begin{equation}
\label{eq_decel}
\Gamma\propto \left\lbrace\begin{array}{l}
\left({E\over n}\right)^{1/8}\,t^{-3/8}\ \ \ {\rm uniform\ medium}\\
\left({E\over A}\right)^{1/4} t^{-1/4}\ \ \ {\rm wind}\\
\end{array}\right.
\end{equation}
then lead to $t_p\propto E_{\gamma}^{1/3}$ and $t_p\propto E_{\gamma}$ in the uniform medium and wind cases respectively, showing a trend 
opposite to the observed one.
\subsection{Reverse shock scenario}
We now suppose that the ejecta emitted by the central engine is made of a ``head'' 
with material at high Lorentz factors ($\Gamma \sim 10^2$ - $10^3$), followed by a ``tail''
where the Lorentz factor decreases to much smaller values, possibly close to unity. The head 
is responsible for the prompt emission while the reverse shock propagating through the tail 
makes the afterglow.

\begin{figure}
\includegraphics[width=0.45\textwidth]{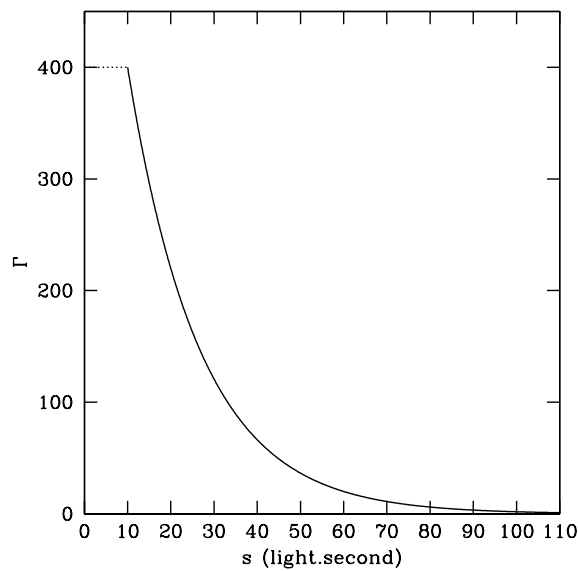} 
\caption{Lorentz factor in the ejecta as a function of the distance from the front (in light.seconds). 
The ``head'' (from 0 to 10 light.seconds) is made of material with typical Lorentz factor $\overline{\Gamma}=400$ while in the 
tail $\Gamma$ decreases from $400$ to unity following \eq{eq_lfDist}, so that ${dE\over d{\rm Log}\,\Gamma}$ is constant. 
}
\label{fig_lf}
\end{figure}

We adopt for the head a constant energy injection rate ${\dot E}_{\rm H}$ 
for a duration of 10 s. We do not specify the distribution of the Lorentz factor 
and simply consider its average value, supposed to be $\overline{\Gamma}=400$. 
The tail that follows lasts for $100$ s but this value is not critical
as long as it remains sufficiently short not to exceed the duration of the early steep decay phase observed at the
beginning of most X-ray light curves. 
We start with a simple case where the distribution of energy in the tail 
${dE\over d{\rm Log}\Gamma}$ is constant from $\Gamma=400$ to $\Gamma=1$. This can be obtained by adopting a constant 
energy injection rate ${\dot E}_{\rm T}$ and 
a Lorentz factor of the form
\begin{equation} 
\label{eq_lfDist}
\Gamma_{\rm T}(s)=400 ^ {  1.1-s/(c\times100{\rm s}) }\ ,
\end{equation}
from $s=10$ to $110$ light.seconds, the distance $s$ being counted from the front to the back of the flow (see \fig{fig_lf}).

Using the methods described in \citet{genet_2007} we have obtained the power $P_{\rm diss}(t)$ dissipated by the reverse shock  
as a function of arrival time to the observer for ${\dot E}_{\rm H}=10{\dot E}_{\rm T}=5\,10^{52}$ erg.s$^{-1}$ (so that equal amounts
of energy are injected in the head and tail) and two possibilities for the burst environment: ({\it i}) a uniform medium 
with $n=1000$ cm$^{-3}$ (supposed to be representative of a massive star environment)  
or ({\it ii}) a stellar wind with a wind parameter $A_*=1$.
Going from the dissipated power to actual light curves depends on the assumptions
that have to be made for the microphysics parameters.
The general shape 
of the early X-ray afterglow light curves however remains globally similar to the evolution of $P_{\rm diss}(t)$ 
so that some conclusions can already be reached without having to consider the uncertain post-shock microphysics.    
%
%

\begin{figure*}
\begin{center}
\begin{tabular}{ccc}
\includegraphics[width=0.45\textwidth]{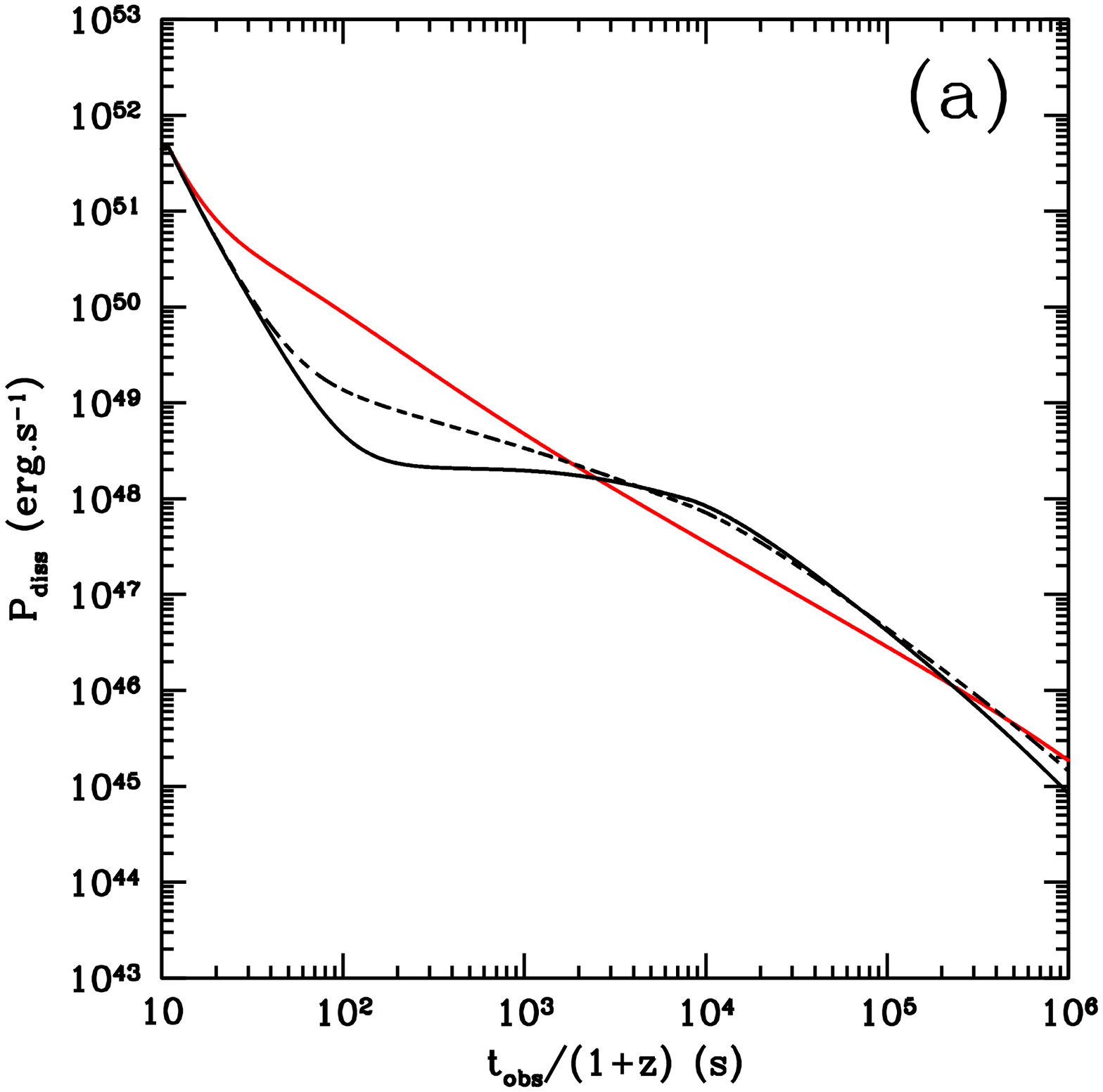} &
\includegraphics[width=0.45\textwidth]{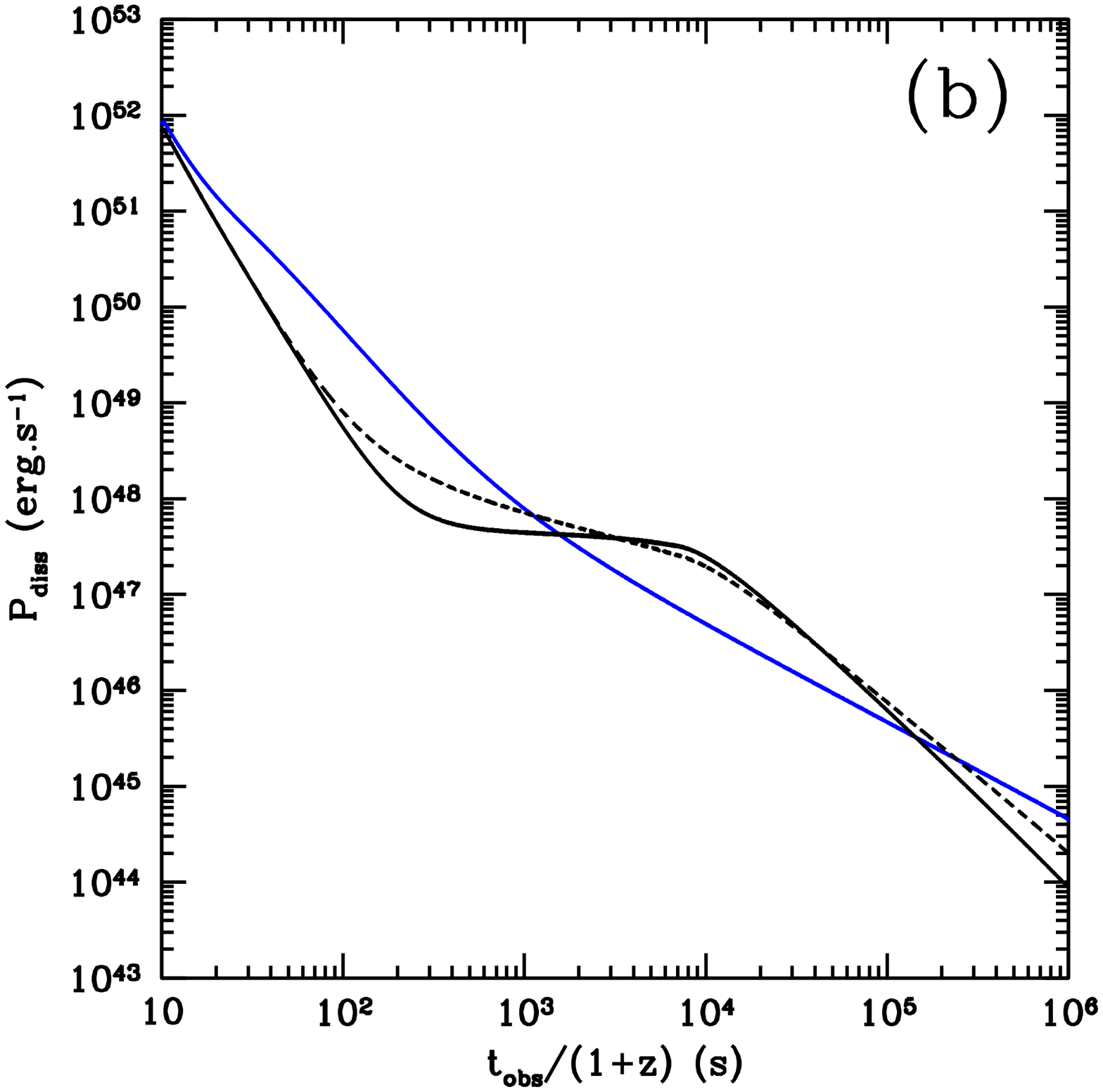}
\end{tabular}
\end{center}
\caption{Dissipated power in the reverse shock as a function of observer time for equal amounts of energy $E_{\rm H}=E_{\rm T}=5\,10^{53}$ erg in the head and tail. 
The distribution of energy in the tail as a function of Lorentz factor is given by \eq{eq_lfDistPeak}.
(a): Uniform external medium of density $n=1000$ cm$^{-3}$, $\Gamma_*=12$, $q=q^{\,\prime}=1.5$ (dashed line) and $q=q^{\,\prime}=2.5$ (full line); 
(b): stellar wind with $A_*=1$, $E_{\rm H}=E_{\rm T}=5\,10^{53}$ erg,
$\Gamma_*=20$, $q=q^{\,\prime}=3$ (dashed line) and $q=q^{\,\prime}=4.5$ (full line). 
The red and blue lines have  $q=q^{\,\prime}=0$ and correspond to a uniform distribution of energy ${dE\over d{\rm Log}\Gamma}$ in the tail.}
\label{fig_rsExample}
\end{figure*}

\fig{fig_rsExample} (red and blue curves) shows that if energy is evenly distributed in the tail (constant ${dE\over d{\rm Log}\Gamma}$) 
the dissipated power approximately decays as
$t^{-1}$ after about 1000 s, for both a uniform and a wind ambient medium. 
The contrast $\kappa = \Gamma/\Gamma_{\rm bw}$, where $\Gamma$ and $\Gamma_{\rm bw}$ are respectively the Lorentz factors of the unshocked ejeta and the blastwave,
is larger for the uniform medium than for the wind case ($\kappa\simeq 2$ and $\sqrt{2}$ respectively, see \citealt{genet_2007}).
As seen in \fig{fig_rsExample} the dissipated power is therefore larger (by a factor $3-5$) in the uniform medium. 

We now vary the energy deposition in the tail, concentrating more power at some value of the Lorentz factor.
We have for example considered a simple model where
\begin{equation}
\label{eq_lfDistPeak}
{\dot E}_{\rm T}(\Gamma)= \left\lbrace\begin{array}{cl}
{\dot E}_*\left({\Gamma\over \Gamma_*}\right)^{-q}\ \ \ {\rm for}\ \ \Gamma>\Gamma_*\\
{\dot E}_*\left({\Gamma\over \Gamma_*}\right)^{q^{\,\prime}}\ \ \ {\rm for}\ \ \Gamma<\Gamma_*\\
\end{array}\right.
\end{equation}
the value of ${\dot E}_*$ being fixed by the total energy injected in the tail. 
Figures \ref{fig_rsExample}a and \ref{fig_rsExample}b respectively show the dissipated power for $\Gamma_*=12$, $q=q^{\,\prime}=1.5$ and $2.5$ (uniform medium) and
$\Gamma_*=20$, $q=q^{\,\prime}=3$ and $4.5$ (stellar wind) with $E_{\rm H}=E_{\rm T}$ in both cases. 
When energy deposition is more concentrated (increasing $q$ and $q^{\,\prime}$) a plateau progressively forms and becomes flatter.
The value of $\Gamma_*$ in \eq{eq_lfDistPeak} fixes the duration of the plateau as it  corresponds to the time when the reverse shock
reaches $s_*$ where $\Gamma_{\rm T}(s_*)=\Gamma_*$. The $q$ parameter controls the flatness of the plateau while $q^{\,\prime}$ controls the decay 
index after the plateau. 

The duration $t_{\rm p}$ of the plateau is roughly given by
\begin{equation}
\label{eq_tp}
t_{\rm p}\sim \left\lbrace
\begin{array}{l}
6\times10^5 E_{{\rm H},53}^{1/3} n^{-1/3}\Gamma_{*,1}^{-8/3}\ {\rm s} \\
10^5\,E_{{\rm H},53}\,A_*^{-1}\,\Gamma_{*,1}^{\,-4}\ \ \ {\rm s}\\
\end{array}\right.
\end{equation}
for a uniform and wind medium respectively. 
\eq{eq_tp} corresponds to the situation of a decelerating shell that does not receive any
supply of energy, contrary to the present case where material from the tail is continuously catching up. 
It however remains approximately correct as long as $E_{\rm T}$ does not largely exceeds the energy $E_{\rm H}$ in the head of the ejecta
(as it happens in models where the plateau is made by energy injection into the forward shock discussed in \sect{sect_lateInject}). 

An analytical solution corresponding to the results of \fig{fig_rsExample} can be obtained from the following expression of
$P_{\rm diss}$ (Genet et al, 2007)
\begin{equation}
\label{eq_pdiss}
P_{\rm diss}={dM\over d\Gamma}\,{d\Gamma\over dt}\,\Gamma\,e c^2\ ,
\end{equation}
where $M(\Gamma)$ gives the distribution of mass as a function of the Lorentz factor in the tail, $\Gamma(t)$
is the Lorentz factor of the tail material just being shocked at observer time $t$ (without the $(1+z)$ time dilation factor) and
$e$ is the fraction of the incoming material kinetic energy dissipated in the reverse shock. 
From \eq{eq_lfDistPeak} 
we get
\begin{equation}
\label{eq_dmdg}
{dM\over d\Gamma}={{\dot E}_*\over \Gamma_* c^3}\,\left({\Gamma\over \Gamma_*}\right)^{\pm q-1}{ds\over d\Gamma}=
{{\dot E}_*\,\tau\over \Gamma_* c^2}\left({\Gamma\over \Gamma_*}\right)^{\pm q-1}\times {1\over \Gamma}\ ,
\end{equation}
with $\tau=100/{\rm ln}\,400$ s (we do not distinguish between $q$ and $q^{\,\prime}$ in \eq{eq_dmdg} to simplify the notation).
The total energy in the tail is given by
%
%
\begin{eqnarray}
\label{eq_et}
E_{\rm T} &=& \int_{10}^{110} {\dot E}_{\rm T}\,dt
={\dot E}_*\tau\times\varphi_{qq^{\,\prime}} \, ,
\end{eqnarray}
with
\begin{equation}
\varphi_{qq^{\,\prime}} = {1\over q}+{1\over q^{\,\prime}} \, .
\end{equation}
%
%
%
We now write $\Gamma(t)$ as
\begin{equation}
\label{eq_lft}
\Gamma(t)\simeq\Gamma_*\left({t\over t_{\rm p}}\right)^{-\gamma}\ ,
\end{equation}
with $\gamma=3/8$ (resp. $1/4$) for a uniform medium (resp. a stellar wind) and with $t_{\rm p}$ being
the duration of the plateau. 
Then, combining Eqs (\ref{eq_pdiss}-\ref{eq_dmdg}-\ref{eq_et}-\ref{eq_lft}) and the expression of $e$ 
\begin{equation}
e={1\over 2}\left[1-(1-2\gamma)^{1/2}\right]^2\ ,
\end{equation}
(Genet et al, 2007) we finally obtain
\begin{equation}
\label{eq_pdissFinal}
P_{\rm diss}(t)={E_{\rm T}\over t_{\rm p}\,\varphi_{qq^{\,\prime}}} F(\gamma)\,\left({t\over t_{\rm p}}\right)^{\pm \,q\gamma-1}\ ,
\end{equation}
with
\begin{equation}
F(\gamma)={\gamma\over 2}\left[1-(1-2\gamma)^{1/2}\right]^{2}\ .
\end{equation}
The decay indices before and after the break at the end of the plateau are
\begin{equation}
\label{eq_rel_indices}
\left\lbrace\begin{array}{ll}
\alpha_1=\gamma q-1\\
\alpha_2=-\gamma q^{\,\prime}-1   
\end{array}\right.
\end{equation}
so that a flat plateau is expected for $q=1/\gamma$ (i.e. $q=8/3$ and $4$ in the uniform medium and
wind cases respectively). For the examples shown in \fig{fig_rsExample}, \eq{eq_rel_indices} gives $\alpha_1=-7/16$ and
$-1/16$ for $q=1.5$ and $2.5$ (uniform medium) and $\alpha_1=-1/4$ and $1/8$ for $q=3$ and $4.5$
(wind). 
If we impose a decay index $\alpha_2=-1.5$ after the plateau
we get the condition $q^{\,\prime}=1/2\gamma$ (i.e. $q^{\,\prime}=4/3$
and $2$ for the uniform medium and wind cases respectively). With our simple choice of $q=q^{\,\prime}$ in \fig{fig_rsExample} 
the decay is steeper when the plateau is flatter. 

\section{Building a sequence of models}
\label{sect_sequence}

\subsection{Forward shock scenario}
\label{fs_scenario}

It has been shown in \sect{subsect_fsPrinciple} that a transition in the behavior of $\epsilon_e$ (from rising to constant) at a fixed density $n_0$ marks the end of the
plateau at a time $t_{\rm p}$ given by \eq{eq_tp_n0}. The X-ray luminosity $L_{\rm p}$ at $t=t_{\rm p}$ then writes from Eqs.(3) and (4)
\begin{equation}
L_{\rm p}\propto E^{p+2\over 4}t_{\rm p}^{-{3p-2\over 4}}\propto t_{\rm p}^{-p}\propto E_{\gamma,{\rm iso}}^p
\end{equation}
as long as the microphysics parameters at the end of the plateau and the gamma-ray efficiency do not vary much from burst to burst. 
\fig{fig_sequence}a shows a sequence of afterglow light curves
corresponding to different values of the isotropic gamma-ray energy release and the following choice of parameters: 
$\epsilon_e=0.1\,(n/n_0)^{-1}$ for $n>n_0=15$ cm$^{-3}$ and $\epsilon_e=0.1$ for $n<n_0$, 
$A_*=0.5$, $p=2.2$, $f_{\gamma}=0.2$. 
It was obtained with a detailed calculation 
where the evolution of each elementary shocked shell is considered separately 
\citep{beloborodov_2005} except for the pressure, which is uniform throughout 
the whole shocked ejecta. The electron population and magnetic field of each newly shocked shell are 
computed taking into account the corresponding shock physical conditions 
and microphysics parameters. Then, each electron population is followed 
individually during the whole evolution, starting from the moment of 
injection, and taking into account radiative and adiabatic cooling. 
The resulting light curves somewhat differ from the simple analytical prediction of Sect.4.1. The plateaus do not stay all flat, 
the brightest ones being slowly rising.

\subsection{Reverse shock scenario}
\label{rs_scenario}

\begin{figure*}
\begin{center}
\begin{tabular}{cc}
\includegraphics[width=0.45\textwidth]{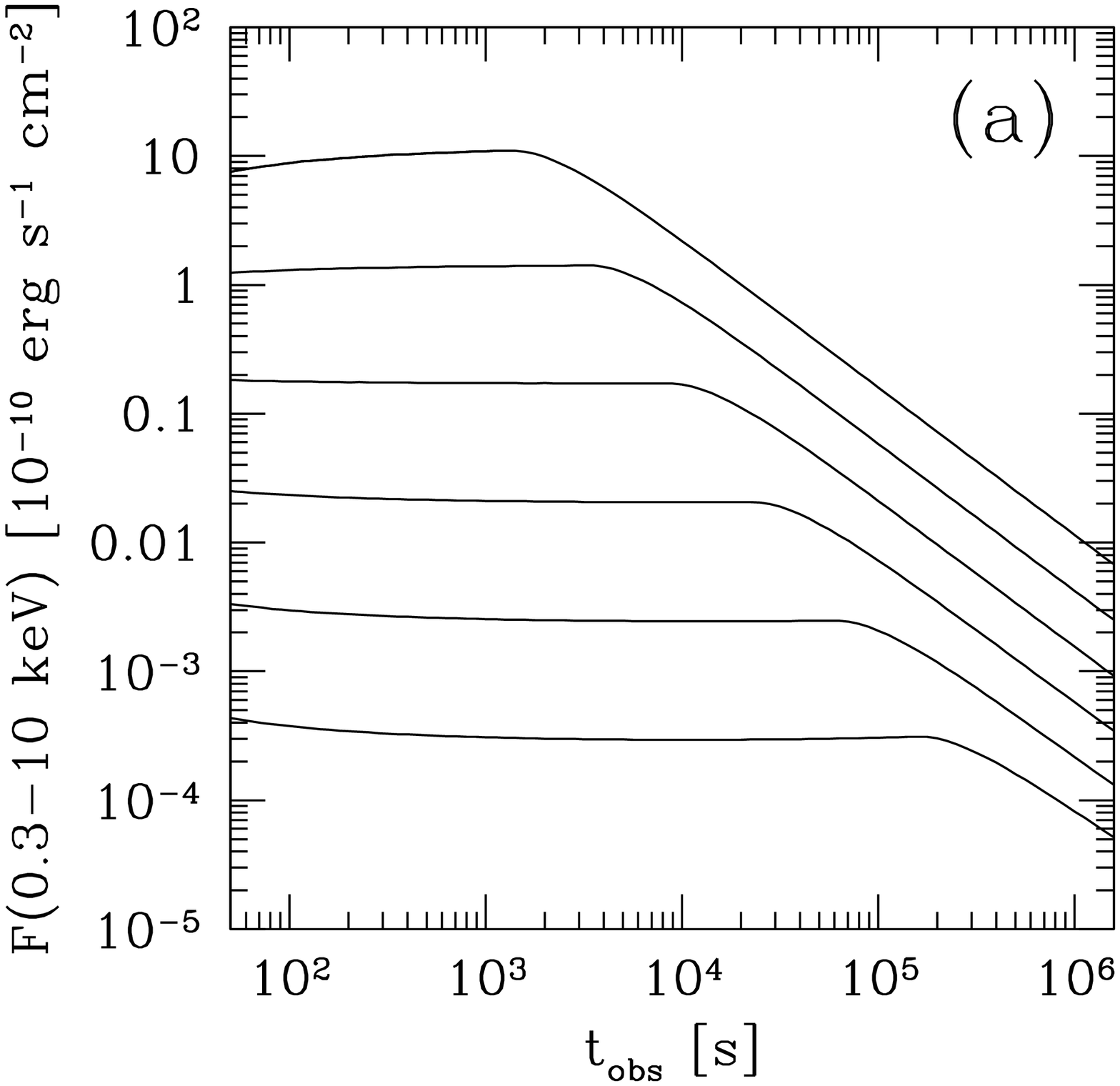} &
\includegraphics[width=0.45\textwidth]{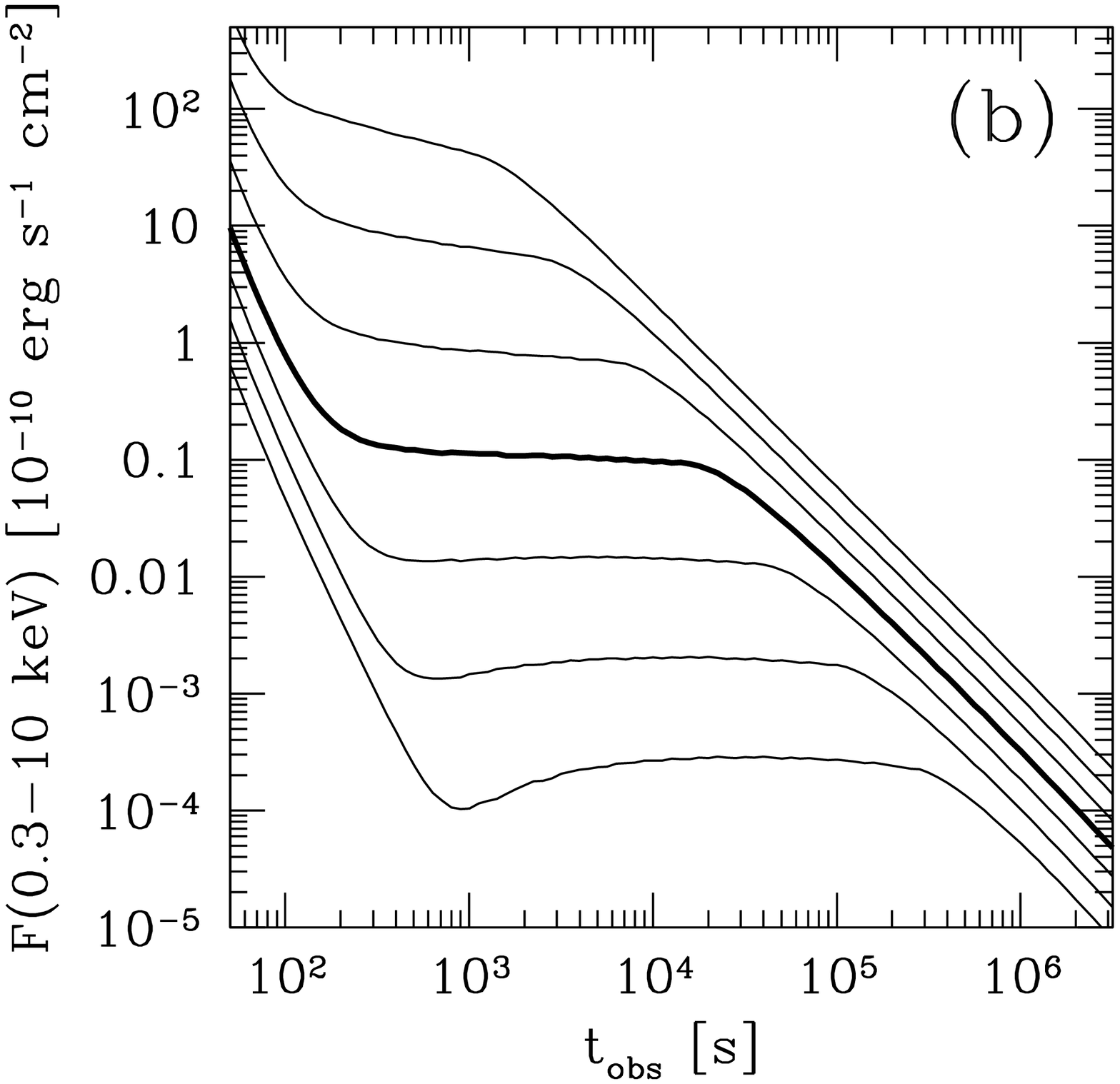}
\end{tabular}
\end{center}
\caption{Sequences of X-ray afterglow light curves with plateaus. (a): forward shock scenario with $\epsilon_e\propto n^{-1}$ for 
$n>n_0=15 \ \mathrm{cm^{-3}}$, a wind parameter $A_*=0.5$ and a gamma-ray efficiency $f_\gamma=0.2$. 
The bottom curve corresponds to an energy injected into the forward shock of $8.5\times10^{51}$ erg and the others by 
successive multiplication of the energy by a factor $F=2.5$;
(b): reverse shock scenario with $\epsilon_e=0.1$,
$\epsilon_B=0.01$, an external medium of uniform density $n=1000$ cm$^{-3}$, a distribution of power in the tail given by \eq{eq_lfDistPeak} with 
$q=8/3$ and $q^{\,\prime}=4/3$. The thick light curve has $E_{\rm H}=E_{\rm T}=2\,10^{54}$ erg and $\Gamma_*=16$, \
while the three others above (resp. below) are obtained by successively multiplying (resp. dividing) the energies by $F=2.5$ and the Lorentz factors by $F^{1/2}$.  
In both panels an index $p=2.2$ for the electron spectrum and a redshift $z=1$ have been assumed.}
\label{fig_sequence}
\end{figure*}

Using \eq{eq_tp} it is possible to link the duration of the plateau to the gamma-ray energy release $E_{\gamma,{\rm iso}}$
if $\Gamma_*$ depends on the burst energy. A relation $\Gamma_{\rm H}\propto E_{\gamma,\rm iso}^{1/2}$ is suggested from the work of 
\citet{liang_2010} and \citet{ghirlanda_2012} based on the rising time of the optical light curve, but \citet{hascoet_2013}
have shown that it partially results from selection effects and has an intrinsic scatter much larger than originally inferred. 
Nevertheless we adopt $\Gamma_*\propto E_{\gamma,\rm iso}^{1/2}$ for simplicity, keeping in mind a potential large dispersion, see \S\ref{corr_scenarios} below.
If moreover the gamma-ray efficiency 
\begin{equation}
f_{\gamma}={E_{\gamma,\rm iso}\over E_{\rm H}}
\end{equation}
does not vary much from burst to burst, we obtain
\begin{equation}
t_{\rm p}\propto E_{\rm H}^{-1}\propto E_{\gamma,{\rm iso}}^{-1} 
\end{equation}
for both a uniform medium and a stellar wind. Together with \eq{eq_pdissFinal}
this fixes the dissipated power during the plateau phase
\begin{equation}
P_{\rm diss}\propto t_{\rm p}^{-2}\propto E_{\gamma,{\rm iso}}^2\ .
\end{equation}
To now compute a sequence of X-ray light curves from the dissipated power we have to fix the microphysics parameters $\epsilon_e$ and
$\epsilon_B$ in the shocked material for which we adopt the fiducial values  $\epsilon_e=0.1$
$\epsilon_B=0.01$. The results for a uniform external medium of density $n=1000$ cm$^{-3}$ are shown in \fig{fig_sequence}b. They were obtained with
the same method of calculation used in the forward shock case and outlined in Sect.5.1.
We start with a model
having $E=E_{\rm H}=E_{\rm T}=2\,10^{54}$ erg, $\Gamma_*=16$, $q=8/3$ and $q^{\,\prime}=4/3$ and then construct the sequence by multiplying or dividing
$E_{\rm H}$ and $E_{\rm T}$ by a the same factor $F$
(i.e. we keep $E_{\rm H}=E_{\rm T}$) and simultaneously $\Gamma_{\rm H}$ and $\Gamma_{\rm T}$ by $F^{1/2}$. 
This prescription corresponds to $\Gamma_* = \Gamma_0 E_{\rm iso , 53}^{1/2}$ with $\Gamma_0 = 35$.
The sequence obtained for a stellar wind 
is similar, but due to the smaller contrast in Lorentz factor at the shock, the plateau flux is about 3 times smaller for the same value of
the injected energy.
\begin{figure*}
\begin{center}
\begin{tabular}{cc}
\includegraphics[width=0.4\textwidth]{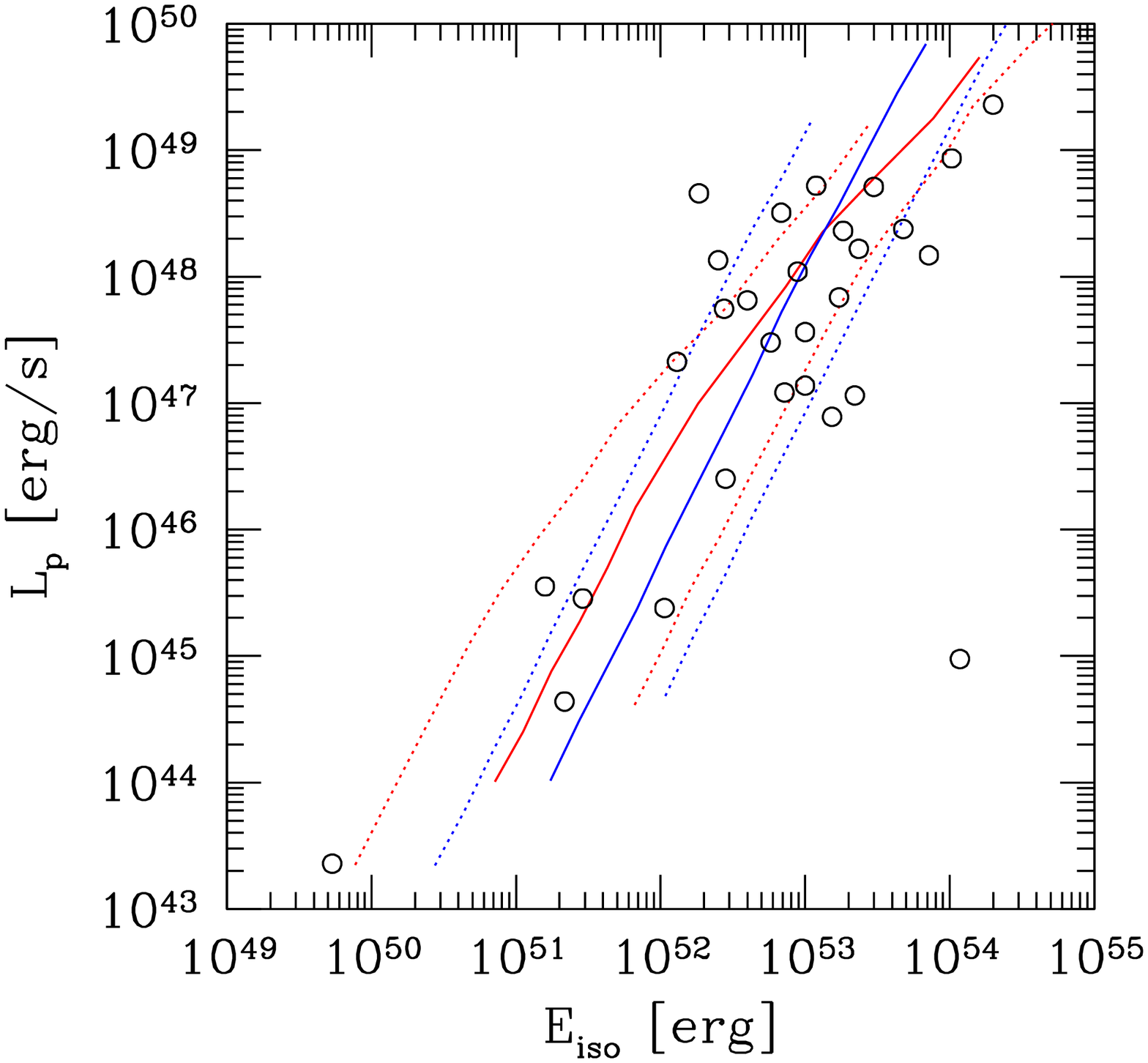} &
\includegraphics[width=0.4\textwidth]{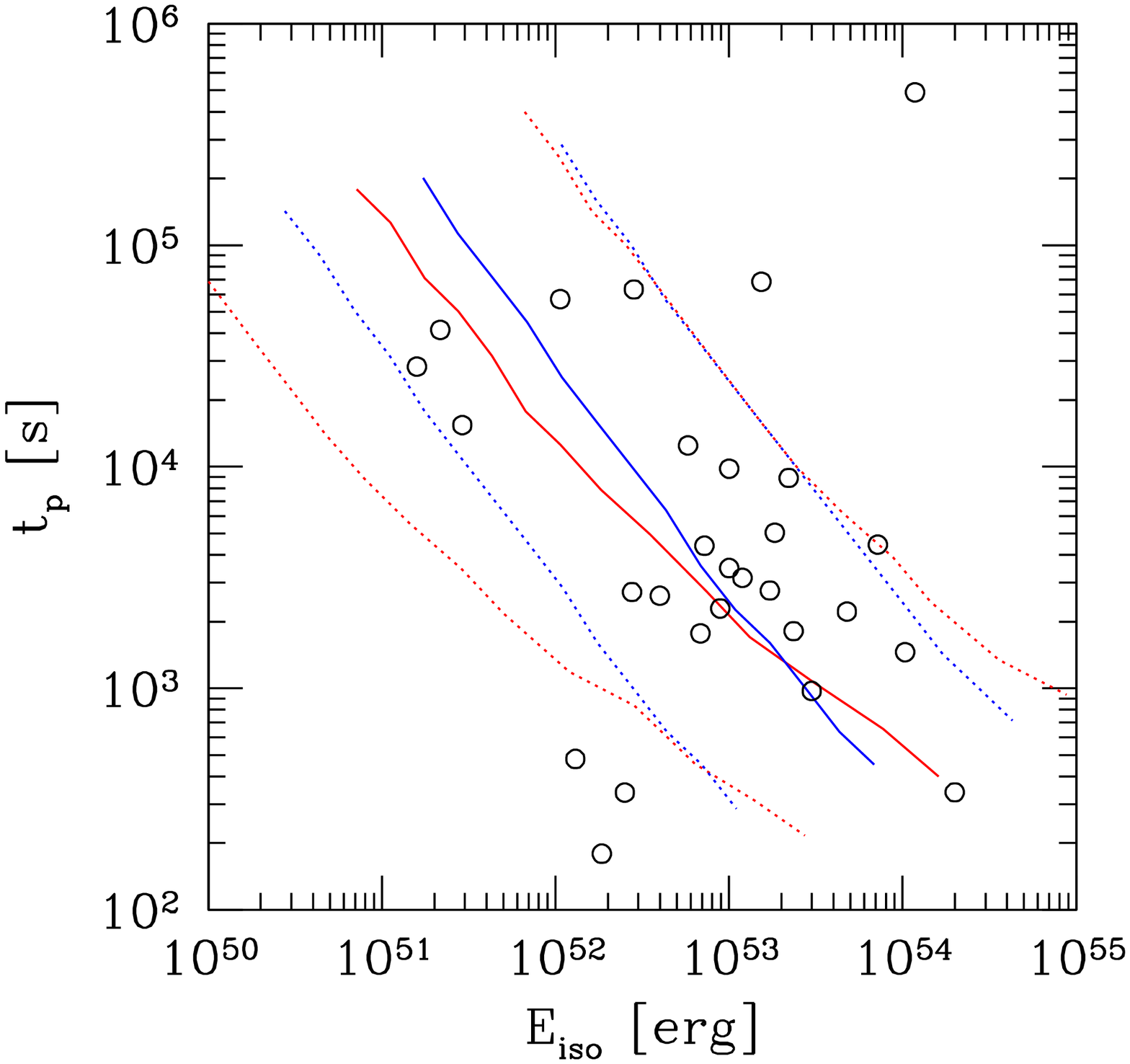} \\
\includegraphics[width=0.4\textwidth]{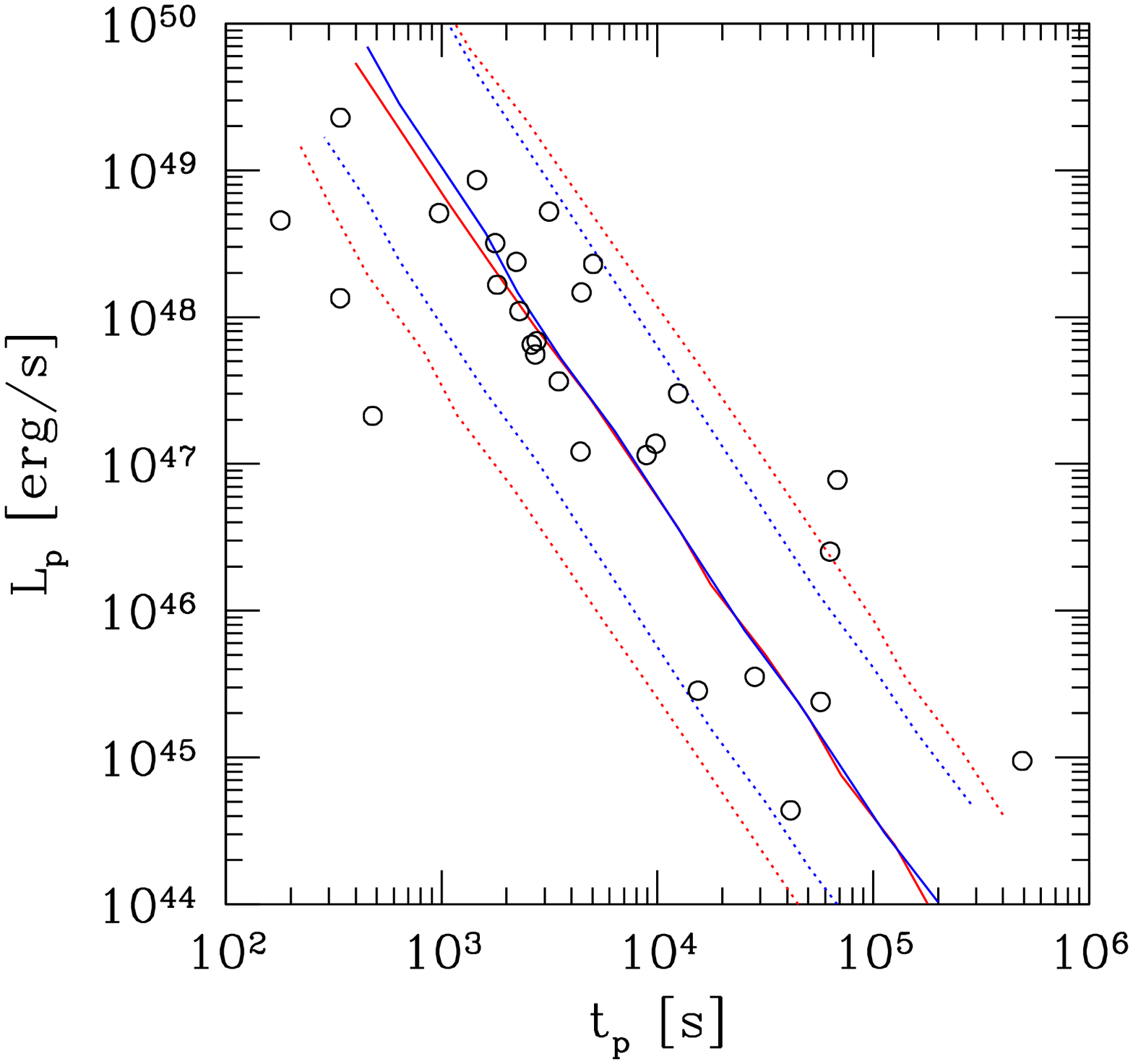} &
\includegraphics[width=0.4\textwidth]{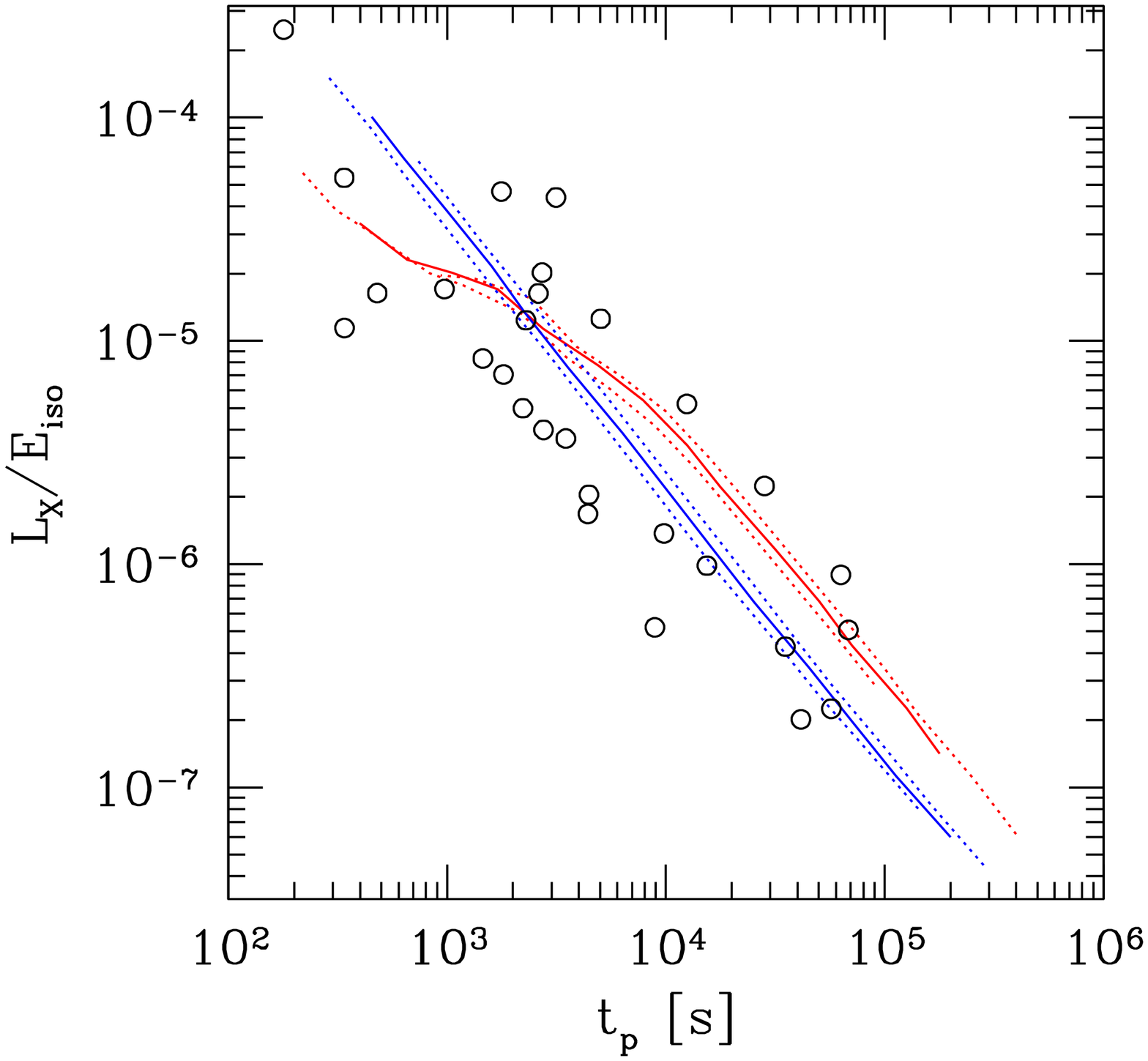} \\
\includegraphics[width=0.4\textwidth]{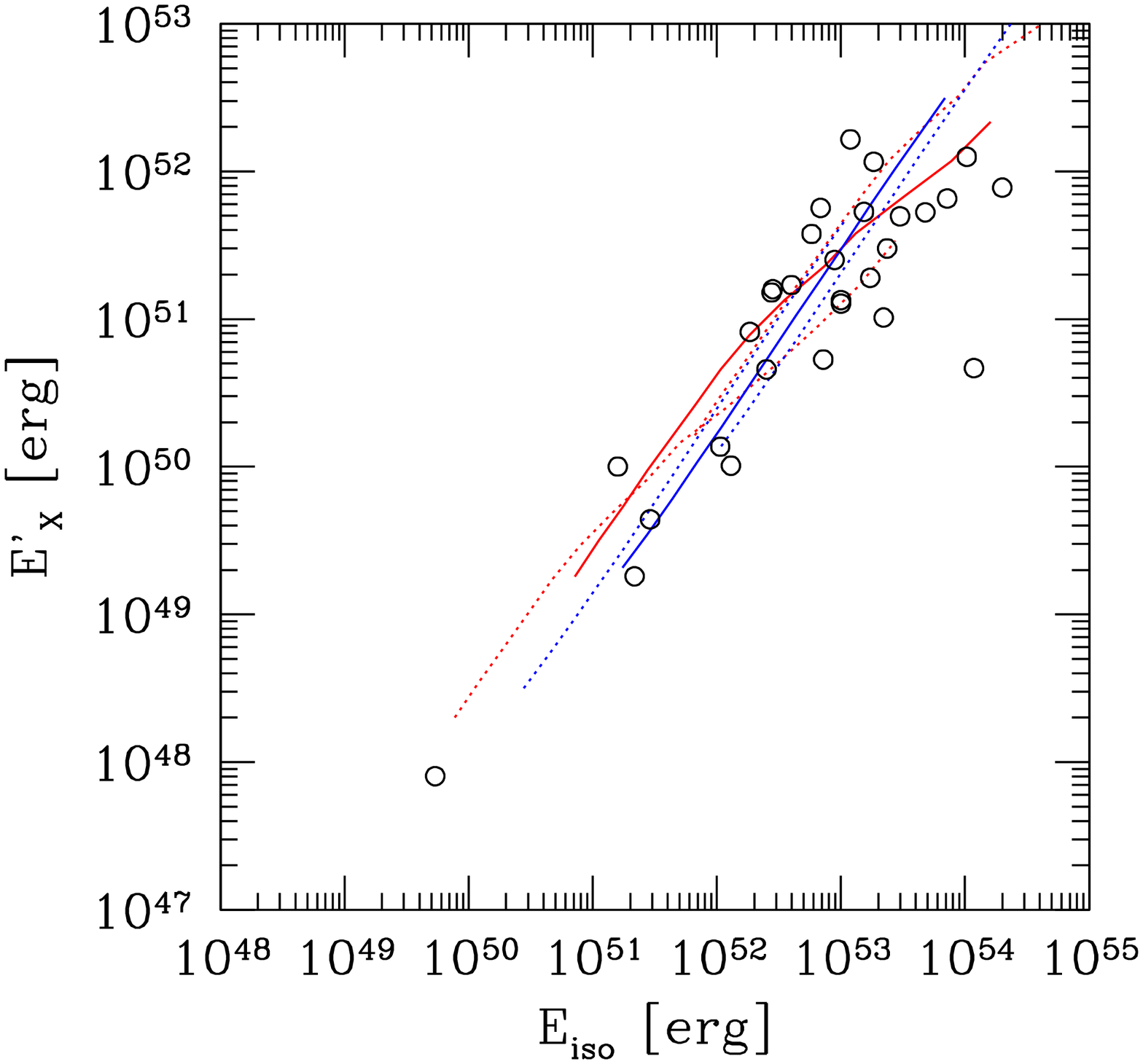} &
\includegraphics[width=0.4\textwidth]{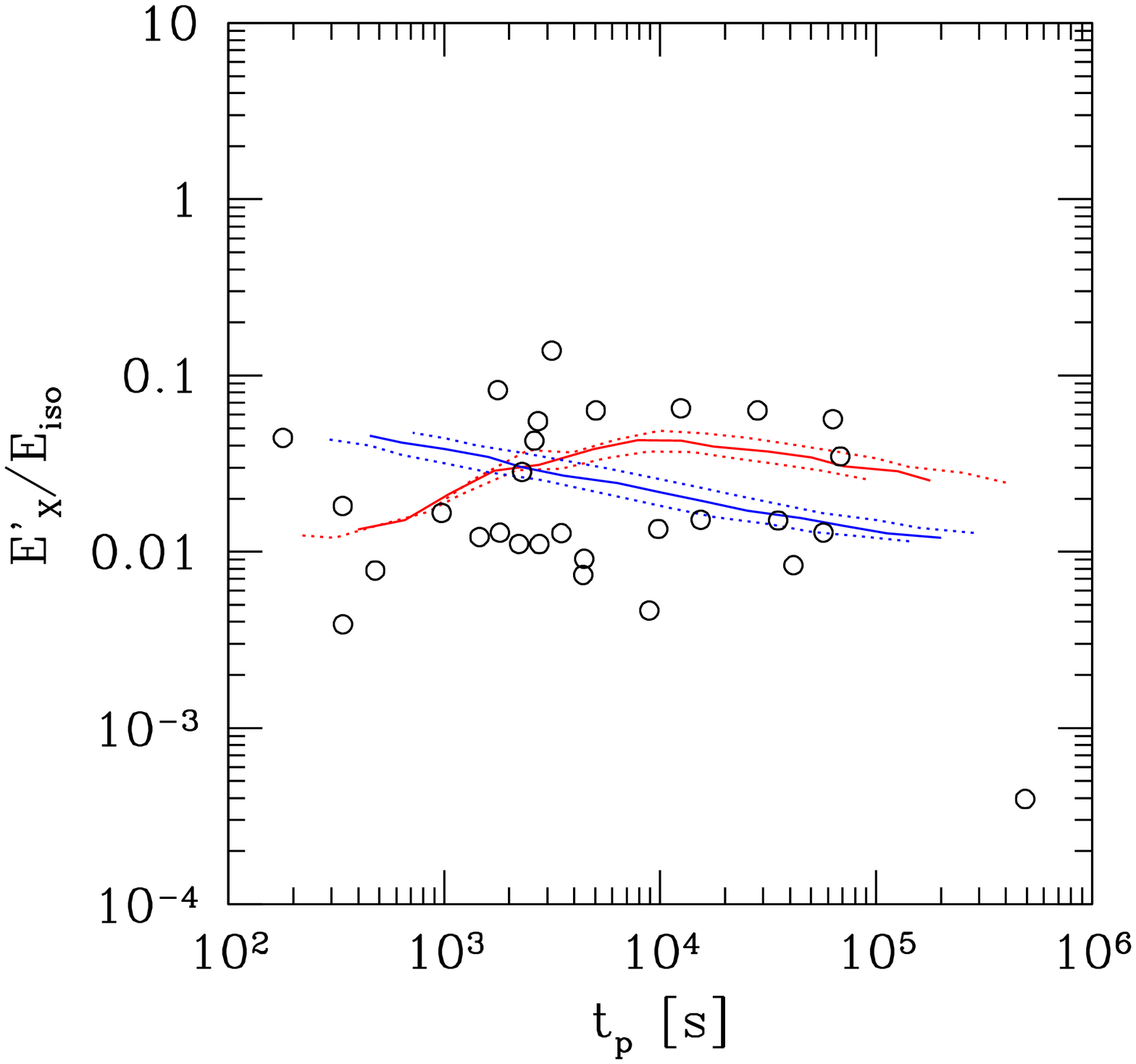}
\end{tabular}
\end{center}
\caption{ 
Prompt-afterglow correlations. 
The blue line correspond to the forward shock case and the red line to the reverse shock case. 
The blue and red dotted lines respectively illustrate the effects of a factor of 3 dispersion in the wind parameter $A_*$ and in the relation $\Gamma_*\propto E_{\gamma,\rm iso}^{1/2}$ (see text for details).
}
\label{fig_correlations}
\end{figure*}
\subsection{Prompt-afterglow correlations}
\label{corr_scenarios}
When the sequences obtained in the previous section are transported back into the burst rest frame, the predicted correlations linking the plateau
duration $t_{\rm p}$, luminosity $L_{\rm p}$, energy release in X-rays $E_X$ 
and the isotropic gamma-ray energy $E_{\gamma,{\rm iso}}$   
can be compared to data. 
This is done in \fig{fig_correlations} for the [$L_{\rm p},E_{\gamma,{\rm iso}}$], [$t_{\rm p},E_{\gamma,{\rm iso}}$], [$L_{\rm p},t_{\rm p}$], [$L_{\rm p}/E_{\gamma,{\rm iso}},t_{\rm p}$], [$E'_X , E_{\gamma,{\rm iso}}$] and [$E'_X/E_{\gamma,{\rm iso}},t_{\rm p}$] relations. 
Since the plateaus in observed bursts are not all flat contrary to our synthetic ones, we have replaced, for a simple comparison between data and models, the true X-ray energy release by the product $E^{\prime}_{\rm X}=L_{\rm p}\times t_{\rm p}$, both for model and data representative points.
%
%
To account for the likely large dispersion of the $\Gamma_* \propto E_{\rm iso , 53}^{1/2}$ relation \citep{hascoet_2013},
we also plot sequences corresponding to $\Gamma_0$ multiplied or divided by $3$. Similarly, in the forward shock scenario we represent sequences where the wind parameter $A_*$ has been multiplied or divided by $3$. 
In some plots this dispersion has little effect, while in some others, especially [$t_{\rm p},E_{\gamma,{\rm iso}}$], 
it is quite large, but still compatible with the scatter of the data.

\section{Discussion and conclusion}
\label{sect_conclusion}

We have addressed in this paper the origin of the plateau phase that is observed in about 50\% of the early afterglow light curves observed by {\it Swift} XRT \citep{nousek_2006}.
We have shown that the commonly invoked cause of plateau formation by continuous energy injection into the forward shock leads to an efficiency crisis for the
prompt mechanism as soon as the plateau duration exceeds $10^3$ seconds.  

We have then discussed two alternatives to energy injection, the first one still in the context of the forward shock scenario, the second in the more
speculative one where the early afterglow is made by a long-lived reverse shock. Within the forward shock scenario a simple way to produce a plateau is to reduce
the radiative efficiency of the shock by acting on the microphysics parameter $\epsilon_e$. For a wind external medium a simple dependence of the form 
$\epsilon_e\propto n^{-1}$ for $n$ larger than a critical density $n_0$ leads to the formation of a plateau approximately satisfying the prompt-afterglow 
correlations. The possibility of such a specific behavior of $\epsilon_e$ remains to be confirmed but it is striking that playing with
only one parameter of the model can account for both the plateau formation and its phenomenology.

In the reverse shock scenario, the shape of the early afterglow is fixed the distribution of injected power ${\dot E}_{\rm T}(\Gamma)$ 
in the low $\Gamma$ tail that is crossed by the shock.
Using simple power laws for ${\dot E}_{\rm T}(\Gamma)$ we have shown that flat plateaus and correct post-plateau decays can be obtained by adjusting the
indices of the power laws. 
In addition, to satisfy the prompt-afterglow 
correlations the typical Lorentz factor of the ejecta should increase with burst energy.
A relation of the form $\Gamma \propto E_{\gamma,{\rm iso}}^{1/2}$, with a large scatter allowed, provides a reasonable fit of the data.  
Since the reverse shock is more efficient in a uniform rather than in a wind external medium, the same plateau luminosity
can be achieved with $3$ times less energy in the tail and we have then only presented results for this former case. 

The reverse shock scenario represents a true change of paradigm compared to the standard viewpoint. 
It has a much larger flexibility
in terms of shapes of afterglow light curves. In addition to the capability to produce a plateau it can also account for various 
accidents such as bumps or steep slopes that are commonly observed \citep{uhm_2012}. 
We have limited the present study to the X-ray light-curves, 
but extending the analysis to the visible/radio domains might help 
to discriminate between the forward and reverse shock scenarii we have considered.

\section*{Acknowledgments}
It is a pleasure to thank Raffaella Margutti who kindly sent us her data on the prompt-afterglow correlations. 
This work has been financially supported by NSF grant AST-1008334 and the Programme National Hautes Energies (PNHE).

\bibliographystyle{mn2e}

\bibliography{main}

\label{firstpage}

\end{document}